\documentclass[aps,10pt,nolongbibliography,prl,tightenlines,twocolumn,twoside,showpacs,superscriptaddress]{revtex4-1}
\usepackage{amsmath}
\usepackage{subfigure,graphicx,color}
\allowdisplaybreaks
\usepackage{amsmath,amssymb,amsfonts}
\usepackage{xcolor}
\usepackage{slashed}
\usepackage{dcolumn}
\usepackage{bm}
\usepackage[utf8]{inputenc}
\usepackage[T1]{fontenc}
\usepackage{mathptmx}
\usepackage{etoolbox}
\usepackage{appendix}

\usepackage[bookmarks,
                bookmarksopen = true,
                bookmarksnumbered = true,
                linktocpage,
                colorlinks = true,
                linkcolor = blue,
                urlcolor  = blue,
                citecolor = blue,
                anchorcolor = green,
                hyperindex = true,
                hyperfigures]
                {hyperref}

\begin{document}

\title{Partonic Critical Opalescence and Its Impact on the Jet Quenching Parameter $\hat{q}$}
\author{Jing Wu}
	\affiliation{School of Physical Science and Technology, Lanzhou University, Lanzhou 730000, China}
	
\author{Shanshan Cao}
\email[Corresponding author, ]{shanshan.cao@sdu.edu.cn}
\affiliation{Institute of Frontier and Interdisciplinary Science,
Shandong University, Qingdao, Shandong, 266237, China}

\author{Feng Li}
	\email[Corresponding author, ]{fengli@lzu.edu.cn}
	\affiliation{School of Physical Science and Technology, Lanzhou University, Lanzhou 730000, China}
	\affiliation{Research Center for Hadron and CSR Physics, Lanzhou University and Institute of Modern Physics of CAS, Lanzhou 730000, China}
	\affiliation{Lanzhou Center for Theoretical Physics, Key Laboratory of Theoretical Physics of Gansu Province, and Frontiers Science Center for Rare Isotopes, Lanzhou University, Lanzhou 730000, China}
	\affiliation{Frontiers Science Center for Rare Isotopes, Lanzhou University}

\date{\today}

\begin{abstract}
Jet quenching parameter $\hat{q}$ is essential for characterizing the interaction strength between jet partons and nuclear matter. Based on the quark-meson (QM) model, we develop a new framework for calculating $\hat{q}$ at finite chemical potentials, in which $\hat{q}$ is related to the spectral function of the chiral order parameter. A meanfield perturbative calculation up to the one-loop order indicates that the momentum broadening of jets is enhanced at both high temperature and high chemical potential, and approximately proportional to the parton number density in the partonic phase. We further investigate the behavior of $\hat{q}$ in the vicinity of the critical endpoint (CEP) by coupling our calculation with a recently developed equation of state that includes a CEP in the universality class of the Ising model, from which we discover the partonic critical opalescence (PCO) -- the divergence of scattering rate of jets and their momentum broadening at the CEP, contributed by scatterings via the $\sigma$ exchange process. Hence, for the first time, jet quenching is connected with the search of CEP.
\end{abstract}

\maketitle

\section {Introduction}

Imaging the QCD phase structure is one of the central goals of the high-energy nuclear collision programs at the Relativistic Heavy-Ion Collider (RHIC) and the Large Hardon Collider (LHC). It is now generally accepted that a strongly coupled Quark-Gluon Plasma (QGP) is produced when heavy nuclei collide with ultra-relativistic energies, as revealed by the collectivity of soft hadron emission and the jet quenching phenomenon~\cite{Gyulassy:2004zy,Jacobs:2004qv}. While the transition from the QGP to the hadronic matter is known to be a smooth crossover~\cite{Aoki:2006we,Bazavov:2011nk} at vanishing baryon chemical potential ($\mu_\mathrm{B}$), whether there exists a first order phase transition at large $\mu_\mathrm{B}$, and if so, where the critical endpoint (CEP) locates in the QCD phase diagram are still open questions.

Extensive efforts have been devoted to seeking and locating CEP. It has been proposed that QCD criticality can be characterized by the high order cumulants of conserved quantities like baryon number~\cite{Stephanov:2008qz,Asakawa:2009aj,Athanasiou:2010kw,Stephanov:2011pb}. Sophisticated dynamic model calculations~\cite{Jiang:2015hri,Herold:2016uvv,Bluhm:2016byc,Stephanov:2017ghc,Wu:2018twy,Nahrgang:2018afz,An:2019csj,Pradeep:2022mkf}  have been developed to explore signatures of CEP that survive the evolution of the complex system. Experimentally, a deviation of the ratio between the fourth and the second cumulants of the net proton number from unity at certain $\sqrt{s_{\rm NN}}$, plausibly caused  by criticality, has been suggested by the Beam Energy Scan (BES) program at RHIC~\cite{STAR:2021iop}, which waits for confirmation with more precise data soon. Meanwhile, the amplification of the light nuclei multiplicity ratio is recently proposed as an alternative signal of CEP~\cite{Sun:2020zxy}.  

Besides the above observables, jets have served as a successful tool in discovering and probing the QGP. Recent experimental and theoretical advances broaden jet studies from medium modification on jets~\cite{Majumder:2010qh,Armesto:2011ht,Bass:2008rv,JET:2013cls} to jet-induced medium excitations~\cite{Qin:2015srf,Cao:2020wlm}, allowing one to explore jet-medium interactions with jet-correlated soft hadrons, e.g., jet substructures~\cite{Chien:2015hda,Casalderrey-Solana:2016jvj,Tachibana:2017syd,KunnawalkamElayavalli:2017hxo,Luo:2018pto,Chang:2017gkt,Mehtar-Tani:2016aco,Milhano:2017nzm,Chen:2020tbl}, $\gamma$/$Z$-triggered hadron distributions~\cite{Qin:2009bk,Chen:2017zte,Zhang:2018urd,Chen:2020tbl,Yang:2021qtl} and hadron chemistry within and around the jet cone~\cite{Chen:2021rrp,Luo:2021voy}. However, rare discussions are available on utilizing jets to probe the QCD phase diagram. Unlike soft hadrons and light nuclei that carry information at either chemical or kinetic freeze-out, energetic jets interact locally with the QGP, thus encode direct information on the phase boundary and could be a more advanced probe of the critical behavior. In fact, the nuclear modification of hard hadrons has already been measured in the BES program~\cite{STAR:2017ieb}, which shows a hint of additional suppression at certain centrality in Au+Au collisions at 14.5~AGeV, but has been overlooked for years. 

Jet evolution can be characterized by the jet quenching parameter $\hat{q}$~\cite{Baier:2002tc,Majumder:2008zg} that measures its transverse momentum broadening square per unit time. Evaluating $\hat{q}$ from the first-principle method~\cite{Majumder:2012sh,Panero:2013pla,Kumar:2020wvb} or extracting it from the jet  data~\cite{JET:2013cls,JETSCAPE:2021ehl,Xie:2022ght} are ongoing efforts at the frontier of heavy-ion physics. In this work, we use two methods to calculate the jet parameters at finite temperature and chemical potential. Using a quark-meson (QM) model~\cite{Schaefer:2007pw} possessing a chiral phase transition, we calculate $\hat{q}$ at finite $\mu_\mathrm{B}$ up to the one-loop order.
This calculation is further coupled with an Equation of State (EoS) incorporating a CEP in the universality class of the Ising model~\cite{Parotto:2018pwx,Martinez:2019bsn},
from which a prominent enhancement around CEP is found for $\hat{q}$. Such an enhancement, named as ``partonic critical opalescence" (PCO) 
in analogy to the normal critical opalescence, might lead to the additional suppression of the hard hadron yield mentioned above, thus provides a new method for exploring the QCD phase diagram around CEP.

\section{Jet quenching parameter and field-field correlation } 
We consider an approximately on-shell jet quark with momentum $q$ scatters elastically with a medium ($M$) by exchanging momentum $k$. The color-isospin-spin averaged matrix element squared reads~\cite{Majumder:2012sh}:
\begin{align}
W(k)=&\frac{1}{2N_c N_f}\left|\left\langle q+k;X\left|\mathcal{T} e^{-i\int_{0}^{t} dt'H_\mathrm{I}(t')}\right|q;M\right \rangle\right|^2,
\label{eq:matrixW}
\end{align}
in which $t$ denotes the propagation time of the quark inside the medium, $H_\mathrm{I}$ is the interacting Hamiltonian in the interaction picture, $X$ represents the final state of the medium, and $\mathcal{T}$ is the time ordering operator. Here, the initial and final states are assumed to be direct products of the medium and the free quark states, where the latter satisfies the box normalization condition, ${q(x)|q \rangle = e^{iqx}u(q)/\sqrt{2E_{q}V}}$ with $q(x)$ standing for the quark field and $u(q)$ denoting the Dirac spinor. The initial and final jet quark momenta 
are assumed light-like, i.e., $q^2=(q+k)^2=0$. The jet quenching parameter and the scattering rate can then be obtained as
\begin{equation}
    \begin{aligned}
    \hat{q}\equiv&\frac{\langle\vec{k}^{2}_{\perp}\rangle}{t}=\sum_{k,X}\vec{k}^{2}_{\perp}\frac{W(k)}{t}.
\end{aligned}
\end{equation}

In this work, the jet-medium interaction is described using the QM model~\cite{Schaefer:2007pw}, where interactions among quarks are mediated by the mesonic fields, such as $\sigma$ and $\pi$, rather than gluons. In the hadronic phase, these fields can be directly viewed as $\sigma$ and $\pi$ particles, while in the QGP phase, they are only effective descriptions of the force mediator. The QM model is widely employed as a possible imitation of QCD at finite chemical potentials with a chiral phase transition and a corresponding CEP embedded. Although the applicability of QM model at very high temperature might be questionable, we consider it a reasonable description around the phase transition boundary, especially when  exploring the behavior of $\hat{q}$ near the CEP. 
The full Lagrangian of the QM model will be given in the next section, while its interaction sector is
\begin{equation}
H_\mathrm{I}=g\int d^{3}x \bar{q}(\sigma^\prime+i\gamma_{5}\vec{\pi}\cdot\vec{\tau})q,
\end{equation} 
where $\tau$ denotes the Pauli matrix, and $\sigma^\prime $ is the fluctuation of the $\sigma$ field around its expectation value $\sigma^\ast$ given by the gap equation~\cite{Schaefer:2007pw}.

After expanding Eq.~(\ref{eq:matrixW}) in terms of $H_\mathrm{I}$ and keeping only the leading order, one obtains $\hat{q}=\hat{q}_{\sigma}+\hat{q}_{\pi}$ with
\begin{equation}
    \begin{aligned}
    \hat{q}_{\sigma/\pi}&=\frac{g^2}{(2\pi)^{3}N_c N_f}\int \frac{d^{3}\mathbf k}{ E_q E_{q+k}}k^2_{\perp}q\cdot(q+k)\tilde{G}_{\sigma/\pi}(k),
\label{eq:qhatSeparate}
\end{aligned}
\end{equation}
in which
\begin{eqnarray}
\widetilde{G}_{\sigma}(k)&=&\int d^4 x\langle M|\sigma^\prime(0)\sigma^\prime(x)|M \rangle e^{ik\cdot x},\nonumber\\
\widetilde{G}_{\pi}(k)&=&\int d^4 x\langle M|\vec{\pi}(0)\cdot\vec{\pi}(x)|M \rangle e^{ik\cdot x}
\end{eqnarray}
represent the correlation functions of $\sigma$ and ${\pi}$ fields between different locations. The physical significance of the $\sigma$ field in the QM model is the quark condensate, which is considered the order parameter of the chiral phase transition. Therefore, via Eq. (\ref{eq:qhatSeparate}), we connect the jet quenching parameter $\hat q$ with the correlation function of the order parameter, and hence the correlation length that diverges at CEP.

The correlation functions can be further expressed using the spectral functions ($\widetilde{\rho}_{\sigma/\pi}$) via the Kubo-Martin-Schwinger (KMS) relation~\cite{Kubo:1957mj,Martin:1959jp}, 
\begin{equation}
        \widetilde{G}_{\sigma/\pi}(k) =\frac{d_{\sigma/\pi}}{ e^{\beta k_0}-1}\widetilde{\rho}_{\sigma/\pi}(k),
\label{eq:G}
\end{equation}
with $d_{\sigma}=1$ and $d_{\pi}=3$ representing the degeneracy of the $\sigma$ and $\pi$ fields respectively.

\section{Jet quenching parameter at finite chemical potential}
The spectral function in Eq.~(\ref{eq:G}) is evaluated perturbatively within the QM model whose Lagrangian is
\begin{align}
    \mathcal{L}=&\;\bar{q}[i\partial\cdot\gamma-g(\sigma+i\gamma_5{\vec{\pi}}\cdot{\vec{\tau}})]q+\frac{1}{2}(\partial_{\mu}\sigma)^2\nonumber\\
    &+\frac{1}{2}(\partial_{\mu}\vec{\pi})^2-\left[\frac{\lambda}{4}(\sigma^2+\vec{\pi}^2-v^2)^2-c\sigma\right],
    \label{eq:qm_Lagrangian}
\end{align}
where $c$ is the explicit symmetry breaking parameter and ${\lambda}$ is the quartic coupling constant. The parameters are determined by fitting the vacuum $\pi$ and $\sigma$ masses and the pion decay constant as $g=3.2$, $c=1.77\times10^6$~MeV$^3$, $\lambda=19.7$ and $v=87.6$~MeV~\cite{Schaefer:2007pw}. With this setup, the CEP is obtained from solving $(\partial p/\partial n_B)_T=(\partial^2 p/\partial n_B^2)_T=0$ under the mean-field approximation as $T_\mathrm{c}\approx 89~$MeV and $n_B\approx 0.21~$fm$^{-3}$ (or $\mu_\mathrm{c}\approx228~$MeV). Thanks to the presence of the explicit symmetry breaking term ($c\sigma$), the $\sigma$ field is of a non-vanishing expectation value which should be evaluated by minimizing the thermodynamic potential (or solving the gap equation equivalently)~\cite{Schaefer:2007pw}. Therefore, the $\sigma$ field can be decomposed as $\sigma=\sigma^\ast+\sigma^\prime$, where the space-time independent expectation value $\sigma^\ast$ contributes to the effective mass of the medium quark as $M_q = g \sigma^\ast$, while the fluctuating sector $\sigma^\prime$ carries the dynamics of the $\sigma$ field. Hence, the spectral function of the $\sigma$ field actually describes the excitation property of $\sigma^\prime$. 

The spectral functions of $\sigma$ and $\pi$ fields in the QM model can be generally expressed as
\begin{equation}
\widetilde{\rho}_{\sigma/\vec{\pi}}(k)=2\mathrm{Im}\frac{1}{-k_\ast^2+m^2_{\sigma/\pi}+g^2 \Pi_{\sigma/\pi}(k_\ast)},
\label{eq:spectralFnc}
\end{equation}
where $k_\ast \equiv (k^0+i 0^+,\mathbf k)$, $m_{\sigma}^2=\lambda(3\sigma^{*2}-v^2)$ and
$m_{\pi}^2=\lambda(\sigma^{*2}-v^2)$ are the vacuum masses of the $\sigma/\pi$ fields, and
\begin{align}
&\Pi_{\sigma}(i\omega_n,\mathbf k)=\int^{\beta}_0  d^4 x\,\mathrm{tr}[S(x)S(-x)]e^{i(\omega_n t-\mathbf k\cdot\mathbf x)},\nonumber\\
&\Pi_{\pi}(i\omega_n,\mathbf k)=-\int^{\beta}_0  d^4 x\, \mathrm{tr}[S(x)\gamma_5 S(-x)\gamma_5]e^{i(\omega_n t-\mathbf k\cdot\mathbf x)}
\label{eq:SelfEnergy}
\end{align}
are the self energies of $\sigma/\pi$ evaluated up to one fermion loop, with
\begin{equation}
\widetilde{S}(i\nu_n,\mathbf p)=\frac{(i\nu_n -\mu)\gamma_0 -\boldsymbol\gamma\cdot\mathbf p-M_q}{(i\nu_n -\mu)^2 -\mathbf p^2-M^2_q}
\label{eq:SQuark}
\end{equation}
representing the Matsubara propagator of the medium quark. Although the frequencies $i\omega_n\equiv 2\pi i nT$, $i\nu_n\equiv i\pi (2n+1) T$ in Eqs.~(\ref{eq:SelfEnergy}) and~ (\ref{eq:SQuark}) are imaginary and discrete, an analytical continuation $i\omega_n \to k^0+ i 0^+$ is applied when evaluating the self energies in Eq.~(\ref{eq:spectralFnc}).

\begin{figure}[tbp!]
    \centering
    \includegraphics[width=0.23\textwidth]{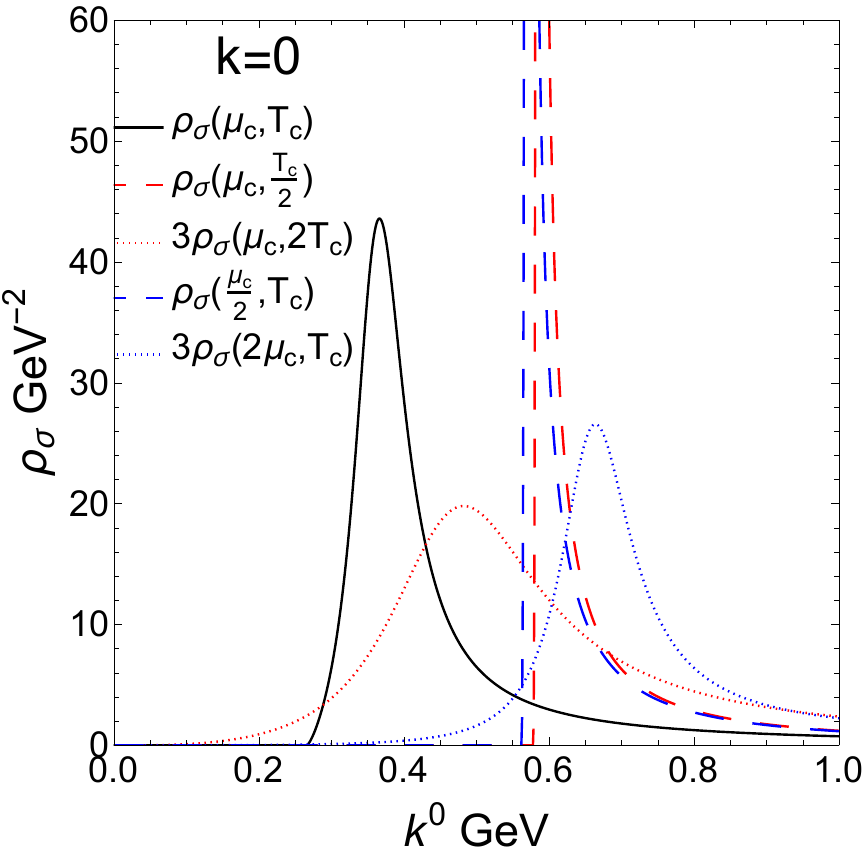} \includegraphics[width=0.23\textwidth]{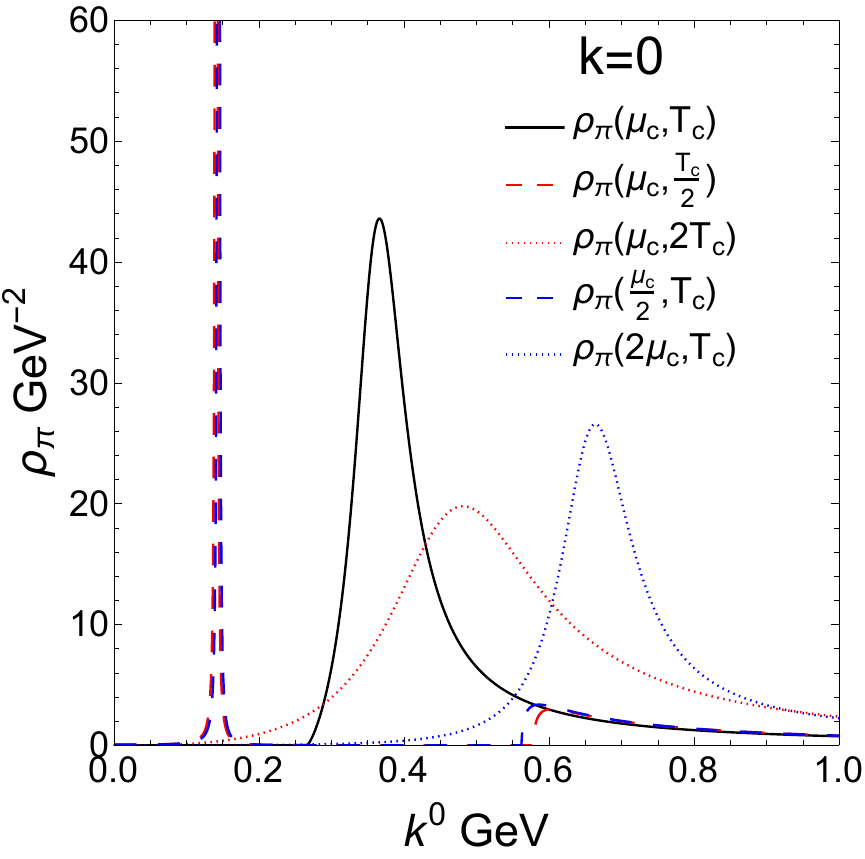}
    \includegraphics[width=0.23\textwidth]{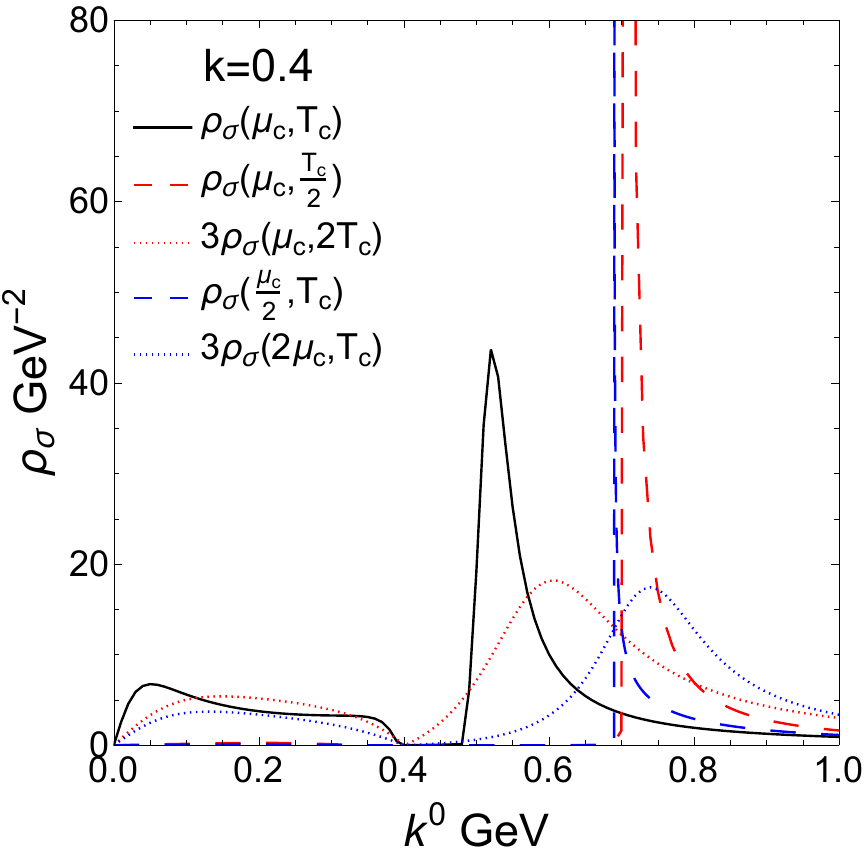}
    \includegraphics[width=0.23\textwidth]{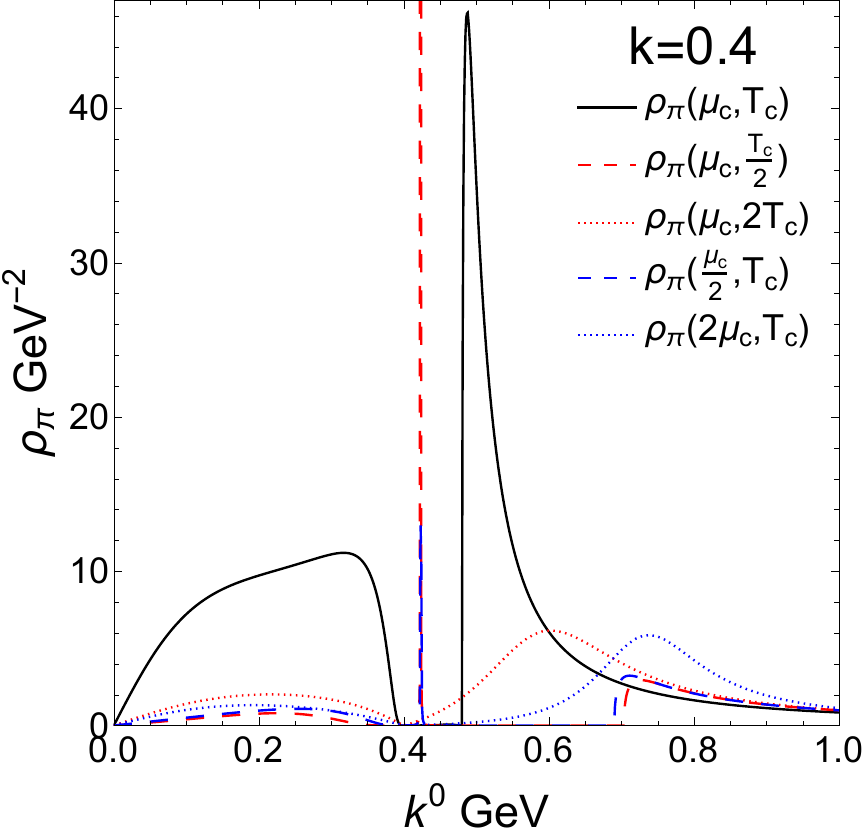}
    \caption{(Color online) The spectral functions of $\sigma$ (left column) and $\pi$ (right column) fields at various temperatures and chemical potentials with $\mathbf k=0$ (upper row) and $|\mathbf k|=0.4$ GeV (lower row), evaluated perturbatively up to the one-loop level.}
    \label{fig:spectralF}
\end{figure}

Shown in the upper panels of Fig.~\ref{fig:spectralF} are the spectral functions of $\sigma$ (left) and $\pi$ (right) fields at $\mathbf k=0$. Those obtained at CEP, in the hadronic (or chiral symmetry breaking) phase and in the QGP (or the chiral symmetry restored) phase are plotted by the solid, dashed and dotted curves, respectively. It should be remarked that the $\sigma$ effective mass, approximately indicated by the peak location of $\rho_\sigma$ at $\mathbf k=0$, first decreases and then increases as the temperature or the chemical potential increases, and achieves its minimum value at CEP (or on the phase boundary), while the $\pi$ effective mass keeps growing with temperature and chemical potential. Meanwhile, the spectral functions of $\sigma$ and $\pi$ fields are almost degenerate in the QGP phase, which is regarded as one feature of chiral symmetry restoration. 

The spectral functions 
at finite momentum transfer ($|\mathbf k|=0.4$~GeV) are presented in the lower panels of Fig. \ref{fig:spectralF}. Besides the similar peak structures shown in the region of $k^0>|\mathbf k|$, additional bumps appear at $k^0<|\mathbf k|$. Since the momentum transfer carried by the mesonic fields is space-like in elastic scatterings, the jet quenching parameter $\hat{q}$ is directly contributed by these bumps. Furthermore, these bumps are drastically suppressed in the hadronic phase, suggesting an almost vanishing $\hat q$ in the hadronic matter.

\begin{figure}[tbp!]
    \centering
    \includegraphics[width=0.23\textwidth]{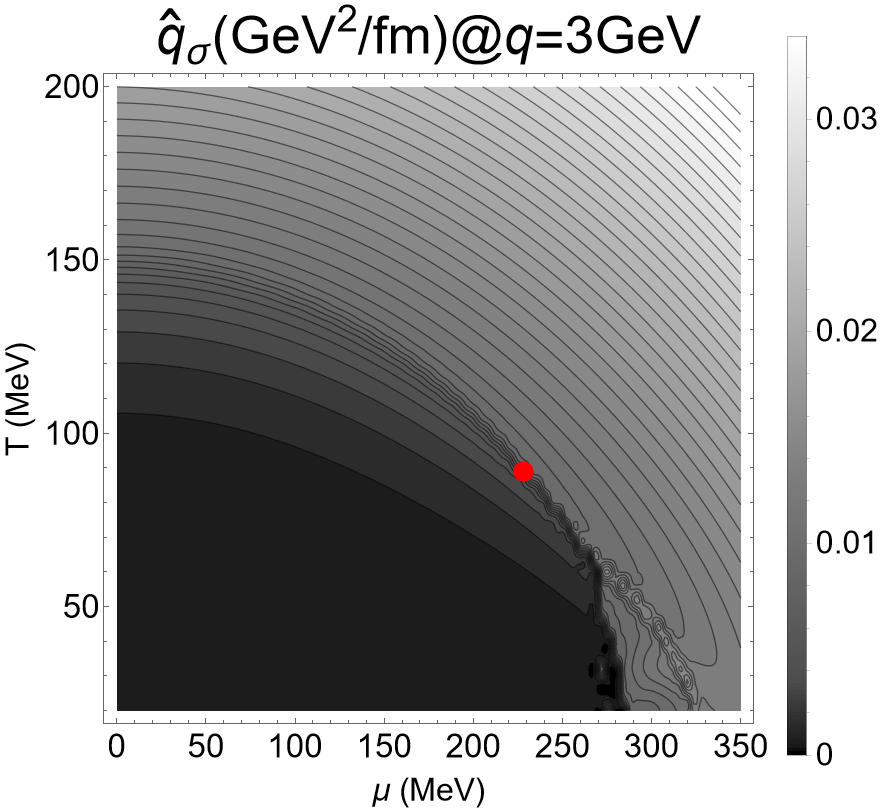}
    \includegraphics[width=0.23\textwidth]{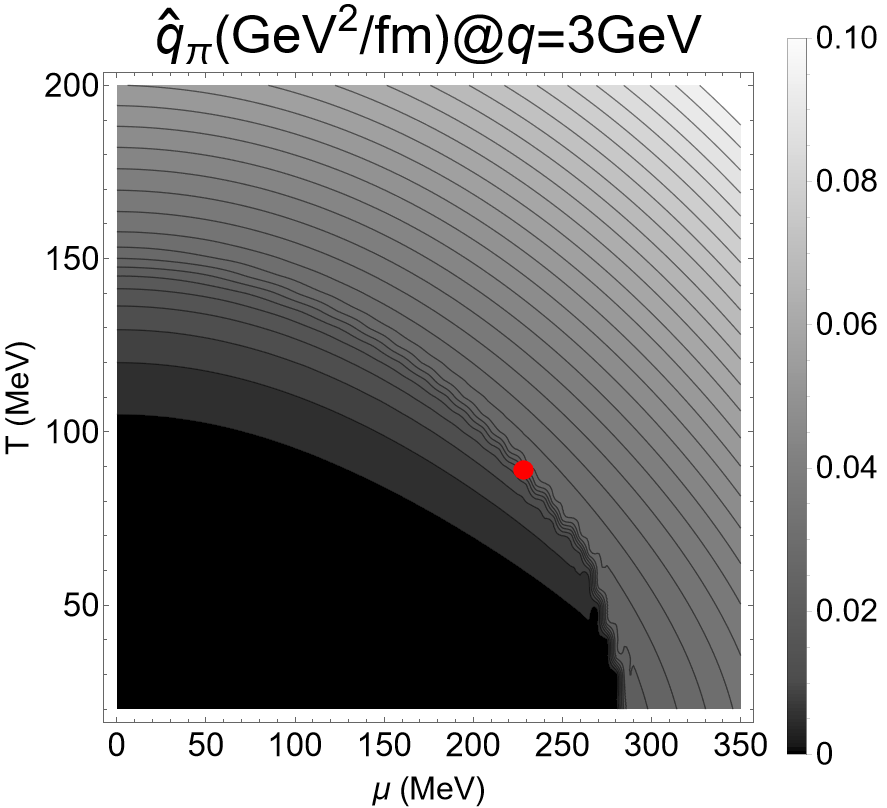}
    \caption{(Color online) Contour plots of the temperature and chemical potential dependent jet quenching parameter $\hat{q}$ for a jet quark at 3~GeV energy, contributed by the $\sigma$ (left panel) and $\pi$ (right panel) exchange processes. The CEP is labeled by the red dot.}
    \label{fig:qhatContour}
\end{figure}
\begin{figure}[tbp!]
    \centering
    \includegraphics[width=0.23\textwidth]{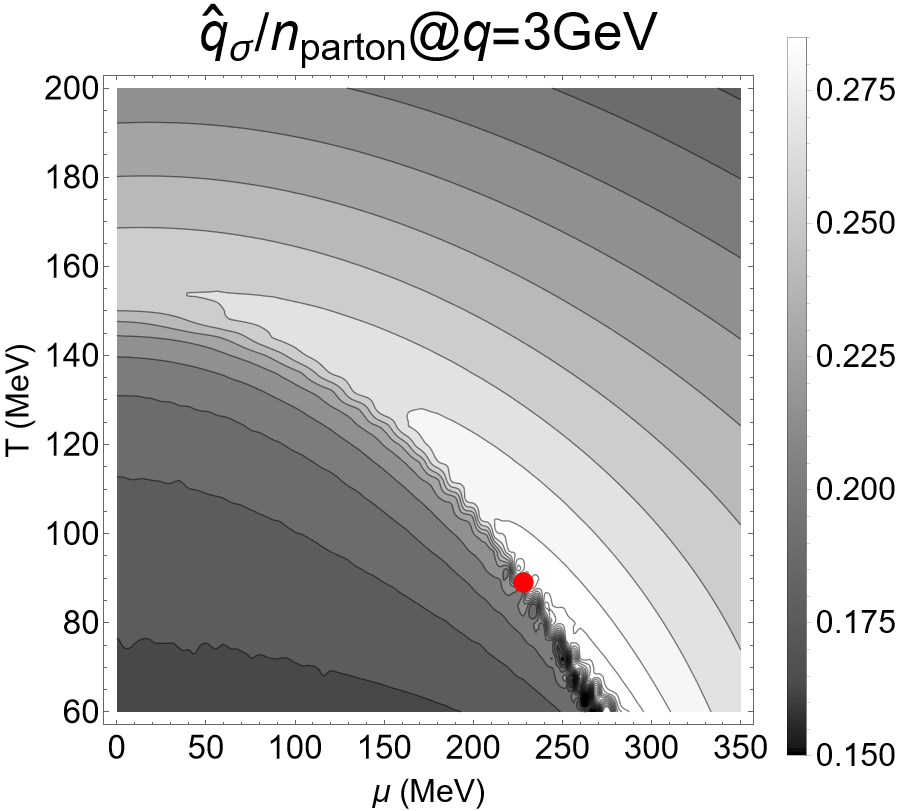}
    \includegraphics[width=0.23\textwidth]{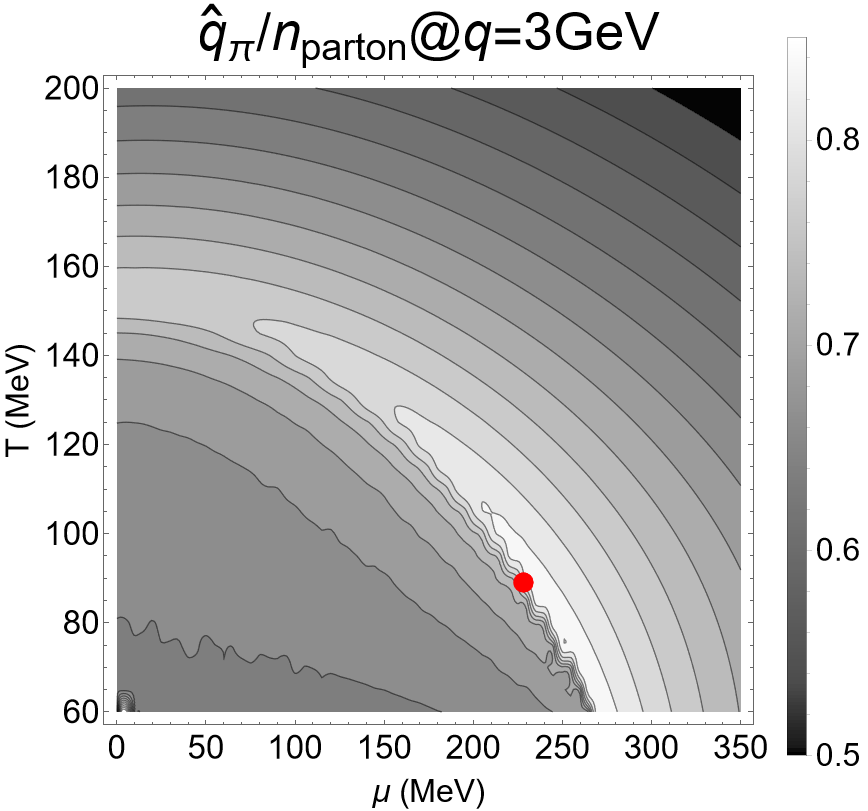}
    \caption{(Color online) Same as Fig.~\ref{fig:qhatContour}, except that $\hat{q}$ is rescaled by the parton number density $n$. 
    }
    \label{fig:qhatnContour}
\end{figure}

By substituting the spectral functions Eq.~(\ref{eq:spectralFnc}) into Eq.~(\ref{eq:G}) and then Eq.~(\ref{eq:qhatSeparate}), we obtain the jet quenching parameter contributed by the $\sigma$ and $\pi$ exchange processes, as shown in the left and right panels of Fig.~\ref{fig:qhatContour} respectively. It is illustrated that $\hat q_{\sigma/\pi}$ almost vanishes in the chiral symmetry breaking phase and grows with not only temperature but also chemical potential. In addition, we find that $\hat{q}$ is approximately proportional to the parton number density $n_{\rm parton}$, which is the summation of the quark and anti-quark number density in the QM model, in the chiral symmetry restored phase.  This is consistent with the empirical scenario that the scattering rate of a jet parton is proportional to the density of scattering centers inside the QGP, as suggested by perturbative QCD calculations~\cite{JET:2013cls}. To further investigate the dimensionless ratio $\hat q / n_{\rm parton}$, we plot them on the phase diagram in Fig.~\ref{fig:qhatnContour} for the $\sigma$ and $\pi$ exchange processes respectively. It shows that quantitatively, $\hat{q}_{\sigma}\sim0.2n_\mathrm{parton}$, $\hat{q}_{\pi}\sim0.6n_\mathrm{parton}$, and thus $\hat{q}\sim0.8n_\mathrm{parton}$ at high temperatures and chemical potentials. Both ratios are enhanced near the phase boundary, especially close to the first-order phase boundary. However, no special behavior of $\hat{q}$ is observed around CEP in either Fig.~\ref{fig:qhatContour} or Fig.~\ref{fig:qhatnContour}. We shall see in the next section that it might be due to the unrealistic CEP obtained under the mean field approach.


\section{Jet quenching parameter near CEP}

The universality of critical behaviors allows one to construct mappings among systems in the same class, hence transfers unknown spectral functions, e.g., those of the quark condensate in QCD, into well-established ones, e.g., those of the order parameter in the Ising model, near CEP. This has recently been developed for dealing with the freezing out procedure near the CEP~\cite{Pradeep:2022mkf} and implemented to study the critical behavior of the bulk viscosity~\cite{Parotto:2018pwx,Martinez:2019bsn}, and is now employed for investigating $\hat{q}$ here, and introduced briefly as follows.

\subsection{Mapping between QCD and Ising model}

In this subsection, we shall denote the densities of the extensive thermodynamic variables of QCD and Ising model as $x^a\equiv(e,n)$ and $x^A\equiv(\epsilon,\psi)$ respectively, where $e$ and $n$ represent the energy and the baryon number densities of the QCD, $\epsilon$ and $\psi$ represent the energy density and the spin density (the order parameter) of the Ising model, respectively. 
The mapping between $x^a$ and $x^A$ is not unique~\cite{2013Phase}, and is usually taken linear in literature~\cite{Parotto:2018pwx,Martinez:2019bsn,Akamatsu:2018vjr} as 
\begin{equation}
\label{eq:Map}
\left[\begin{tabular}{c}
     $\delta e$  \\
     $\delta n $
\end{tabular}\right] = \left[
\begin{tabular}{c c}
    $M_{11} $ &  $M_{12}$\\
    $M_{21}$ & $M_{22}$
\end{tabular}\right]
\left[\begin{tabular}{c}
    $\delta\epsilon$  \\
     $\delta\psi$ 
\end{tabular}\right],
\end{equation}
where $\delta$ represents the variation from the critical value.

The mapping between the thermal conjugated variables, defined as $X_{A/a}\equiv-\partial S / \partial x^{A/a}$ with $S$ being the entropy, is considered linear as well, where $X^a = (-\beta,\beta\mu)$ for QCD, and $X^A = (-\beta^{is},\beta_{c}^{is}H)$ for the Ising model, with $\beta$ and $H$ representing the inverse temperature and the magnetic field respectively. In detail,  it is expressed in~\cite{Parotto:2018pwx,Martinez:2019bsn,Akamatsu:2018vjr} as
\begin{equation}
\label{eq:InverseMap}
\left[\begin{tabular}{c}
     $-\delta\beta$  \\
     $\delta (\beta\mu)$
\end{tabular}\right] =\left[
\begin{tabular}{c c}
    $\bar{M}_{11}$ & $\bar{M}_{12}$\\
    $\bar{M}_{21}$ & $\bar{M}_{22}$
\end{tabular}\right]
\left[\begin{tabular}{c}
    $-\delta\beta_{is}$  \\
     $\delta(\beta_{is} H)$ 
\end{tabular}\right].
\end{equation}

Equations~(\ref{eq:Map}) and~(\ref{eq:InverseMap}) can be further simplified as 
\begin{equation}
\label{eq:SimMap}
    \begin{aligned}
    \left[\begin{tabular}{c}
     $t$  \\
     $\bar{\mu}$
\end{tabular}\right] &=\left[
\begin{tabular}{c c}
    $\Bar{A}_{11}$ & $\bar{A}_{12}$\\
    $\bar{A}_{21}$ & $\bar{A}_{22}$
\end{tabular}\right]
\left[\begin{tabular}{c}
    $r$  \\
     $h$ 
\end{tabular}\right],\\
\left[\begin{tabular}{c}
     $\delta s$  \\
     $\delta n$
\end{tabular}\right] &=\left[
\begin{tabular}{c c}
    $A_{11}$ & $A_{12}$\\
    $A_{21}$ & $A_{22}$
\end{tabular}\right]
\left[\begin{tabular}{c}
    $m$  \\
     $\delta\psi$ 
\end{tabular}\right],
    \end{aligned}
\end{equation}
where $t=\frac{T-T_c}{T_c}$ and $\bar{\mu}=\frac{\mu-\mu_c}{T_{c}}$ are the reduced temperature and chemical potential in QCD, while $r=\frac{T^{Is}-T^{Is}_c}{T^{Is}_c}$ and $h=\frac{H}{T^{Is}_{c}}$ are the reduced temperature and magnetic field in the Ising model, respectively. In addition, $\delta s=\frac{1}{T_c}\delta e-\frac{\mu_c}{T_c}\delta n$ is the variation of entropy density in QCD, and $m\equiv \delta \epsilon / T_c^{is}$.

Universality requires that the singular parts of the entropy are identical for both QCD and the Ising model,  which, as illustrated in Ref.~\cite{Akamatsu:2018vjr}, leads to the relation between the mapping matrices i.e., $M\cdot \bar M^T = A\cdot \bar A^T=I$.

Furthermore, Eq.~(\ref{eq:SimMap}) should map the phase boundary of the Ising model, formulated as $r<0$ and $h=0$, to the phase boundary of QCD, formulated as $\bar\mu > 0$ and $t=t(\bar \mu) \approx k_s \bar\mu$, where $k_s<0$ is the slope of the phase boundary near CEP, which constrains the elements in $A$ and $\bar A$ matrices as $k_s \bar A_{21}=\bar A_{11}$ and $-k_s A_{12}=A_{22}$.

We further require the mapping to be conformal, which results in two more constraints: $-k_s \bar A_{12}= \bar A_{22}$, and $k_s A_{21}= A_{11}$. Finally, to be consistent with Ref.~\cite{Martinez:2019bsn}, we take $\bar A_{12}=1$ and $\bar A_{21}=\frac{2}{3k_s-1}$, and the transforming matrices are

\begin{equation}
    \bar A = \left(
    \begin{tabular}{c c}
      $\frac{2 k_s}{3k_s-1}$   &  $1$\\
      $\frac{2}{3k_s-1}$   &    $k_s$
    \end{tabular}
    \right),~~~  A = \frac{3k_s-1}{1+k_s^2} \left(
    \begin{tabular}{c c}
      $\frac{k_s} 2$   &  $\frac{1}{3k_s-1}$\\
      $\frac 1 2$   &    $-\frac{k_s}{3k_s-1}$
    \end{tabular}
    \right).
\end{equation}

With the above mapping, we can express $\sigma^\prime$ near CEP as $\sigma^\prime=J\cdot A \cdot \phi$, where $J\equiv \left(\left(\partial\sigma / \partial s\right)_n,\left(\partial\sigma / \partial n\right)_s\right)$ is a pair of the thermal derivatives evaluated at CEP, transforming $\sigma'$ into a superposition of $\delta s$ and $\delta n$, and $\phi\equiv(m,\delta \psi)$.

Hence, the correlation and spectral function, i.e., $\tilde G_\sigma$ and $\tilde \rho_\sigma$, in Eq.~(\ref{eq:qhatSeparate}) can be expressed as 
\begin{equation}
\begin{aligned}
G_{\sigma}(x)&\equiv\langle\sigma^\prime(x)\sigma^\prime (0)\rangle=J_l A_{li} \langle\phi^\prime_i(x)\phi^\prime_j(0)\rangle A_{kj}J_k,\\
\widetilde{\rho}_\sigma &=\frac{k^0}{T}\widetilde{G}_{\sigma},
\label{eq:RhoSigmaCEP}
\end{aligned}
\end{equation}
where $\langle\phi^\prime_i(x)\phi^\prime_j(0)\rangle$ can be evaluated non-perturbatively using the Extended Landau-Ginzburg-Wilson (LGW) model as illustrated in detail in the following.

\subsection{Correlations between the Observables in LGW Model}

The LGW Hamiltonian with phonon~\cite{1974Renormalization} is 
\begin{equation}
    \begin{aligned}
    \mathcal{H}=\beta H&=\int d^3 \mathbf{x}\left(\frac{1}{2}(\nabla\psi)^2+\frac{1}{2}\tilde{t}\psi^{2}\right.\\&\left.+\frac{1}{4}u_{0}\psi^{4}+\frac{1}{2C_{0}}m^2+\gamma_{0} m\psi^2\right).
    \label{eq:LGW_H}
\end{aligned}
\end{equation}
where $m$, $\psi$ and $C_0$ are the reduced energy density, order parameter and heat capacity, respectively.

The bare couplings $u_0$, $C_0$, and $\gamma_0$ are corrected in the medium by the thermal loops. The corrected or dressed couplings obey the scaling laws near the CEP as $C\propto\xi^{\frac{\alpha}{\nu}}, \gamma\propto\xi^{\frac{2\beta-1}{\nu}},\widetilde{u}\equiv u-2\gamma^2 C\propto\xi^{2\eta-1} $, where $\xi$ is the correlation length and $\alpha,~\beta,~\delta,~\eta,~\nu$ are the critical indices whose values are taken from Ref.~\cite{ZinnJustinBook} as $\alpha=0.11$, $\beta=0.33$, $\delta=4.78$, $\eta=0.036$, $\nu=0.63$. 

The expectation values $\bar \psi$ and $\bar m$ are obtained by minimizing the Hamiltonian, i.e., $\frac{\delta\mathcal{H}}{\delta m}|_{m=\bar{m}}=\frac{\delta\mathcal{H}}{\delta\psi}|_{\psi=\Bar{\psi}}=0$. The detailed expression for $\bar\psi$ is given in Ref.~\cite{ZinnJustinBook} as $\bar\psi = M_0 R^\beta \theta$, where $R$ and $\theta$ are the functions $r$ and $h$ whose expressions are given later on in Eq.~(\ref{eq:ParameterTransfer}). $\bar\psi$ also obeys the scaling law near the CEP as $\bar\psi\propto \xi^{-\frac{\beta}{\nu}}$. Hence, like the $\sigma$ field, $\psi$ and $m$ fields can also be decomposed as $\psi = \bar \psi + \psi^\prime$ and $m=\bar m+m^\prime$, with $m^\prime$ and $\psi^\prime$ representing fluctuations around their expectation values.

The static correlation functions between $\psi^\prime$ and $m^\prime$ are given non-perturbatively from the partition function as 
\begin{eqnarray}
    \chi_{m} & \equiv & \int d^3 \mathbf x \langle m^\prime(\mathbf x) m^\prime(\mathbf 0)\rangle e^{i \mathbf k \cdot \mathbf x}\nonumber\\
    & = & C + (\gamma C)^2\left[4\Bar{\psi}^2\chi_\psi+4\Bar{\psi} \chi_{\psi:\psi^2}+\chi_{\psi^2}\right],\\
    \chi_{m:\psi} &\equiv& \int d^3 \mathbf x\langle m^\prime(\mathbf x) \psi^\prime(\mathbf 0)\rangle e^{i \mathbf k \cdot \mathbf x}\nonumber\\
    &=&-\gamma C\left[2\Bar{\psi}^2\chi_\psi+\chi_{\psi:\psi^2}\right],\nonumber
\end{eqnarray}
where
\begin{equation}
\chi_{\psi}\equiv \int d^3 \mathbf x \langle \psi^\prime(\mathbf x) \psi^\prime(\mathbf 0)\rangle e^{i \mathbf k\cdot \mathbf x}  = \frac{\xi^2_0({\xi}/{\xi_0})^{2-\eta}}{1+(|\mathbf k|\xi)^{2-\eta}}
\end{equation}
is taken from Ref.~\cite{ZinnJustinBook} with $\xi_0\approx 1
.2 / \sqrt[3]{5}$~fm, satisfying $s_0\xi_0^3\approx 1$, and
\begin{eqnarray}
     \chi_{\psi^2}&\equiv& \int d^3 \mathbf x \langle \psi^{\prime 2}(\mathbf x) \psi^{\prime 2}(\mathbf 0)\rangle e^{i \mathbf k\cdot \mathbf x}\nonumber\\&\approx&2\int\frac{d^3 \mathbf{k^{\prime}}}{(2\pi)^3}\chi_{\psi}(\mathbf{k}-\mathbf{k^\prime})\chi_{\psi}(\mathbf{k^\prime}),\\
     \chi_{\psi:\psi^2}&\equiv& \int d^3 \mathbf x \langle \psi^\prime(\mathbf x) \psi^{\prime 2}(\mathbf 0)\rangle e^{i \mathbf k\cdot \mathbf x}\approx-3\widetilde{u}\Bar{\psi}\chi_{\psi}\chi_{\psi^2}\nonumber
\end{eqnarray}
are evaluated perturbatively.
    
It should be noted that the LGW Model is static, i.e., no time derivative, and hence no time evolution, is involved in Eq. (\ref{eq:LGW_H}). The dynamics of both the $m$ and $\psi$ fields are introduced manually in analogy to Brownian motion as
\begin{equation}
    \begin{aligned}
    \partial_{t}m=-\Gamma_{m}\circ\frac{\delta\mathcal{H}}{\delta m}+\eta_{m},\\
    \partial_{t}\psi=-\Gamma_{\psi}\circ\frac{\delta\mathcal{H}}{\delta\psi}+\eta_{\psi},
    \label{motion equation 2}
\end{aligned}
\end{equation}
which, after linearization and Fourier transformation, are simplified as 
\begin{equation}
    \begin{aligned}
    \widetilde{\Omega}\cdot\left[\begin{tabular}{c}
     $\widetilde{m}^{\prime}$  \\
     $\widetilde{\psi}^{\prime}$
\end{tabular}\right]\equiv\left[\begin{tabular}{c c}
     $-ik^{0}+\frac{\widetilde{\Gamma_{m}}}{C}$ & $2\gamma\Bar{\psi}\widetilde{\Gamma_{m}}$ \\
     $2\gamma\Bar{\psi}\widetilde{\Gamma_{\psi}}$ & $-ik^{0}+\frac{\widetilde{\Gamma_{\psi}}}{\chi_{\psi}}$
\end{tabular}\right]\cdot\left[\begin{tabular}{c}
     $\widetilde{m}^{\prime}$  \\
     $\widetilde{\psi}^{\prime}$
\end{tabular}\right]=\left[\begin{tabular}{c}
     $\widetilde{\eta}_{m}$  \\
     $\widetilde{\eta}_{\psi}$
\end{tabular}\right],
\label{equation 22}
\end{aligned}
\end{equation}
where $\Gamma_{m/\psi}$ and $\eta_{m/\psi}$ are the damping rates and random forces, respectively.

The damping rate of the conserved quantities, such as $\Gamma_m$ is described by Model B ~\cite{Hohenberg:1977ym}, while $\Gamma_\psi$ is described by model H with Kawasaki's approximation~\cite{Hohenberg:1977ym,KAWASAKI19701}, as follows:
\begin{equation}
    \begin{aligned}
    &\widetilde{\Gamma}_{m}=\lambda_m |\mathbf k|^2, \qquad\frac{\widetilde{\Gamma_{\psi}}}{\chi_{\psi}}=\frac{2T_{\rm is}}{3}\left(\frac{\xi_0}{\xi}\right)^3 K(|\mathbf k|\xi)
    ,\\
    &K(x)=\frac{3}{4}[1+x^2+(x^3 +x^{-1})\arctan(x)],
\end{aligned}
\end{equation}
with $\lambda_m=1$~GeV$^{2}$ chosen in this calculation. 

The random forces $\eta$ are of vanishing expectation values but non-vanishing correlation functions, formulated as $\langle\eta_{a}(\vec{x})\eta_{b}(\Vec{y})\rangle=D_{ab}(\Vec{x}-\Vec{y})\delta(x^0-y^0)$, with $a,b$ being $m$ or $\psi$.
$D_{ab}$ is constrained by the fluctuation-dissipation theorem. In detail, the equal-time correlation functions are considered time independent, i.e., $\langle\phi_{i}^\prime(t,\mathbf x)\phi^\prime_{j}(t,\mathbf y)\rangle=\langle\phi^\prime_{i}(t+\delta t,\mathbf x)\phi^\prime_{j}(t+\delta t,\mathbf y)\rangle$, which, combined with the equations of motion Eq.~(\ref{equation 22}), leads to
\begin{equation}
    \begin{aligned}
    \widetilde{D}_{mm}&=2\widetilde{\Gamma}_{m}\left(\frac{\chi_m}{C}+2\gamma\Bar{\psi}\chi_{m:\psi}\right),
   \\
    \widetilde{D}_{m\psi}&=\widetilde {D}_{\psi m}=\frac{\widetilde{\Gamma}_{\psi}}{\chi_\psi}\chi_{m:\psi}+2\gamma \Bar{\psi}\widetilde{\Gamma}_{\psi}\chi_m+\widetilde{\Gamma}_{m}\left(\frac{\chi_{m:\psi}}{C}+2\gamma\Bar{\psi}\chi_\psi\right),\\
    \widetilde{D}_{\psi\psi}&=2\widetilde{\Gamma}_{\psi}+4\gamma\bar{\psi}\widetilde{\Gamma}_{\psi}\chi_{m:\psi}.
    \label{co}
    \end{aligned}
\end{equation}

The correlations between the random forces would generate the correlations between the $m$ and $\psi$ fields according to Eq.~(\ref{equation 22}) as
\begin{equation}
    \begin{aligned}
\widetilde{\langle\phi^\prime_{i}\phi^\prime_{j}\rangle}=\left(\widetilde{\Omega}^{-1}\cdot\widetilde{D}\cdot\widetilde{\Omega}^{\dagger -1}\right)_{ij}.
\label{correlation phi}
    \end{aligned}
\end{equation}
Substituting the obtained $\widetilde{\langle\phi^\prime_{i}\phi^\prime_{j}\rangle}$ in Eq.~(\ref{eq:RhoSigmaCEP}) and then in Eq.~(\ref{eq:qhatSeparate}), we finally obtain the jet quenching parameter near the CEP contributed by the $\sigma$ exchange channel.

\subsection{Numerical Results}

\begin{figure}[tbp!]
    \centering
    \includegraphics[width=0.235\textwidth]{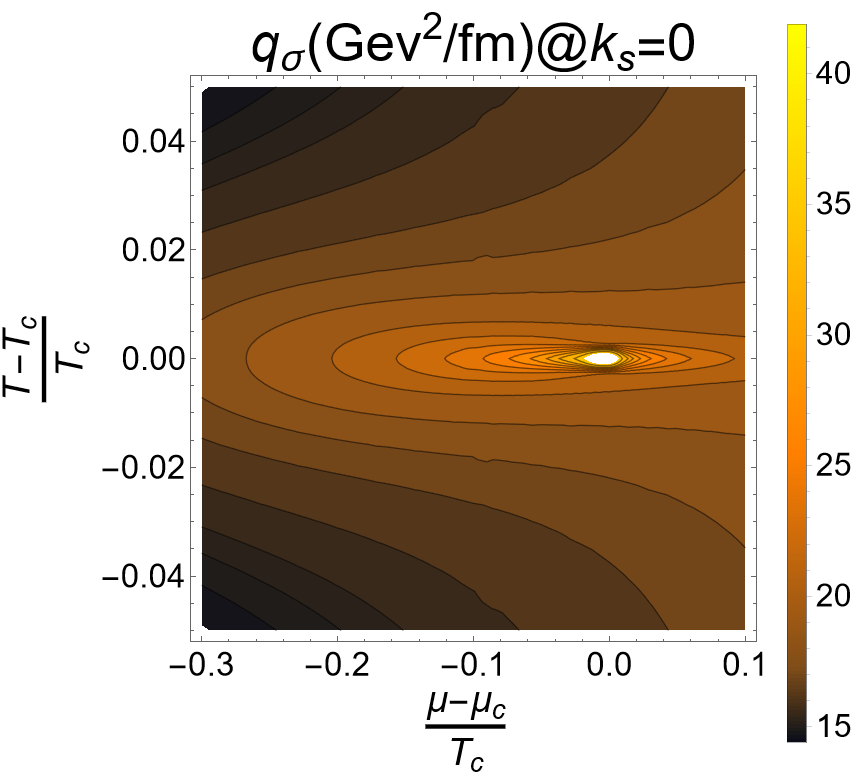}
    \includegraphics[width=0.235\textwidth]{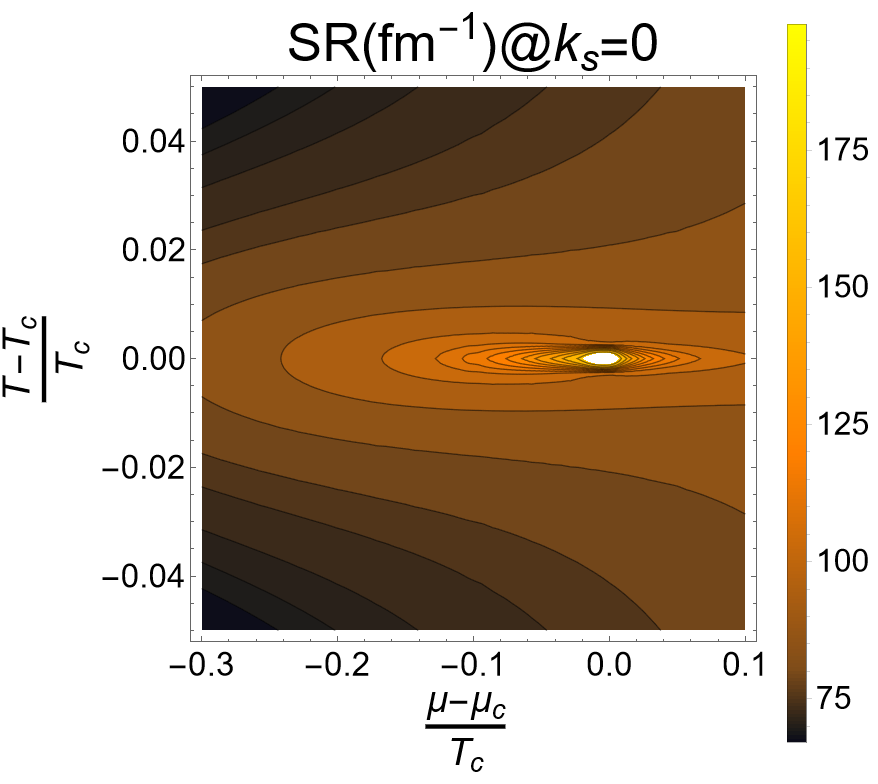}
    \caption{(Color online) The temperature and chemical potential dependencies of $\hat{q}_{\sigma}$ (left) and $SR$ (right) near the CEP, with $k_s=0$.}
    \label{fig:qhat and SR k0}
\end{figure}
\begin{figure}[tbp!]
    \centering
    \includegraphics[width=0.235\textwidth]{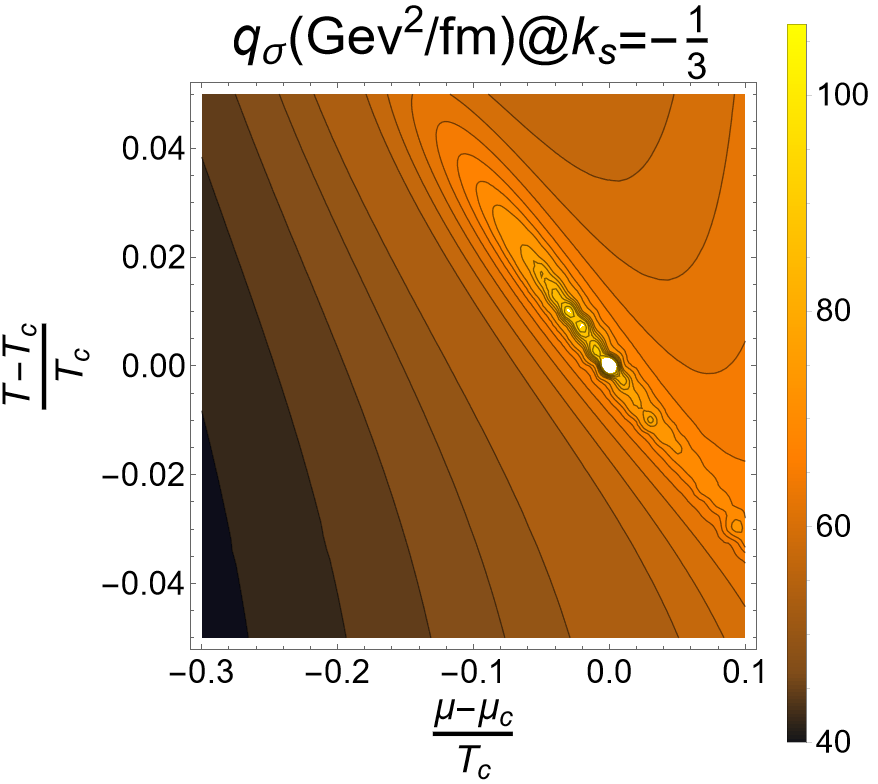}
    \includegraphics[width=0.235\textwidth]{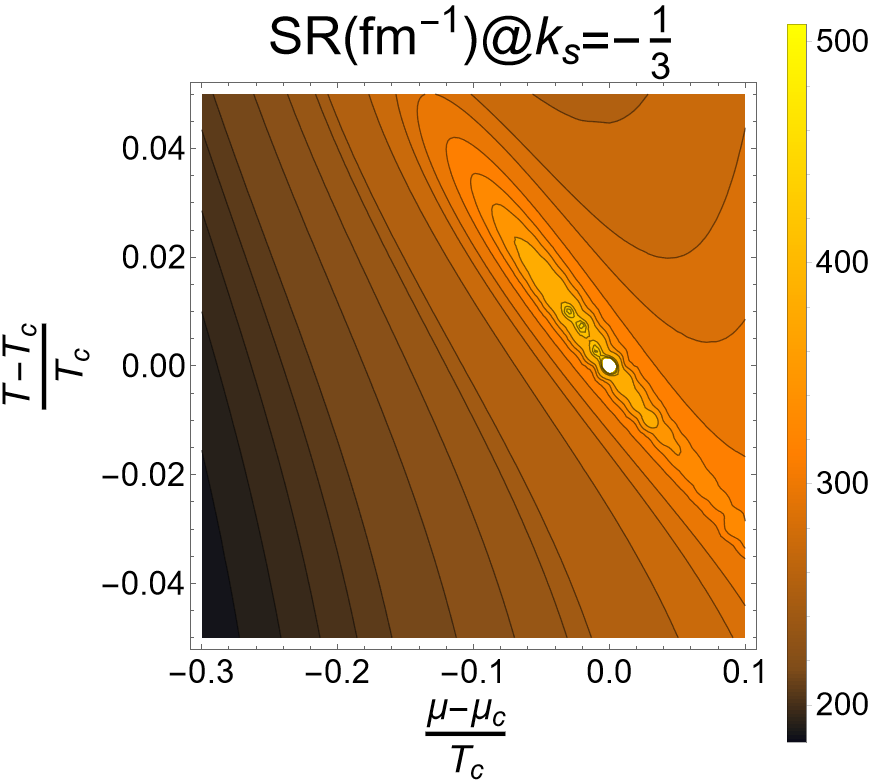}
    \caption{(Color online) The same as Fig.~\ref{fig:qhat and SR k0}, except with $k_s=-\frac{1}{3}$.}
    \label{fig:qhat and SR k3}
\end{figure}
\begin{figure}[tbp!]
    \centering
    \includegraphics[width=0.235\textwidth]{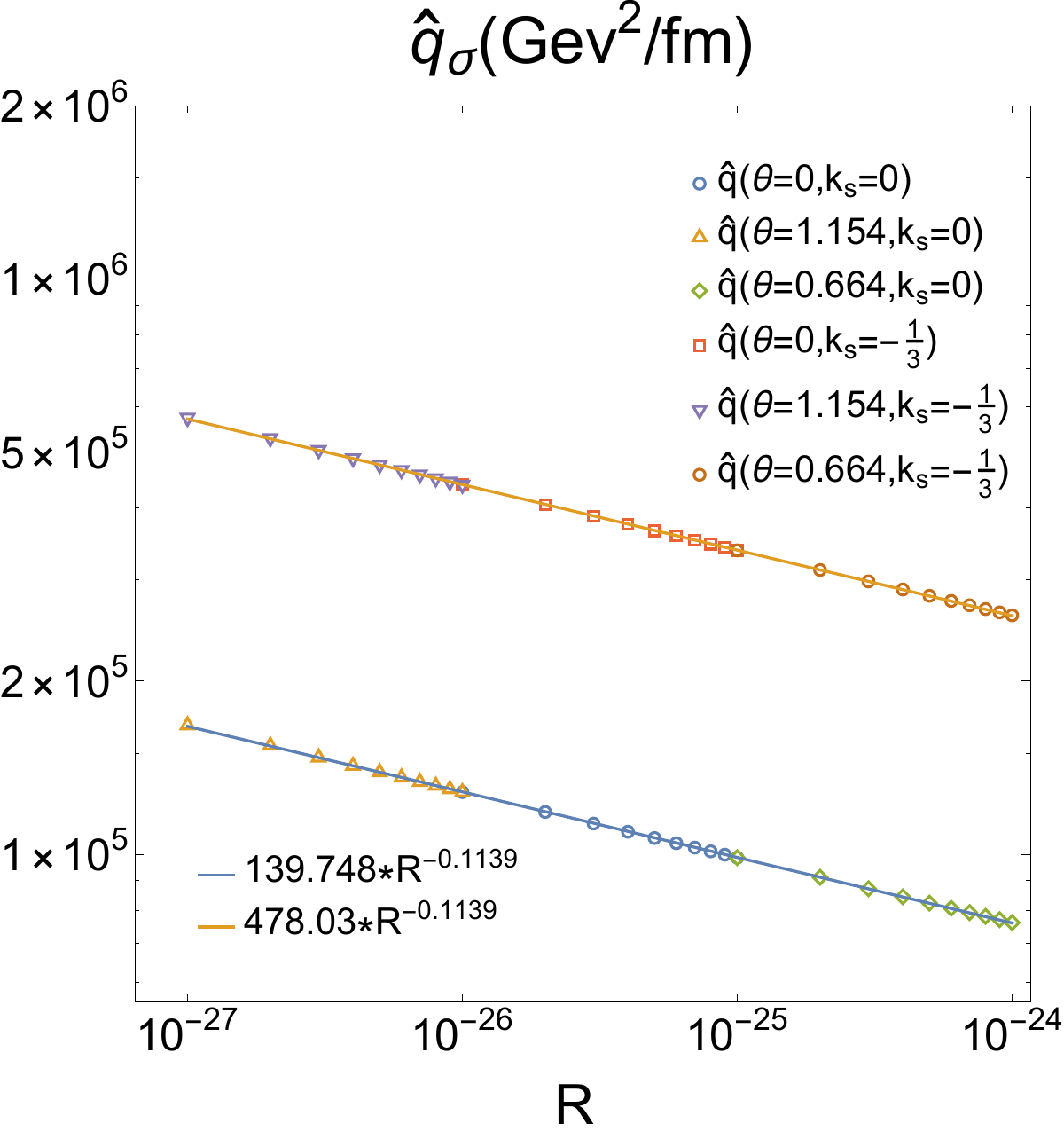}
    \includegraphics[width=0.235\textwidth]{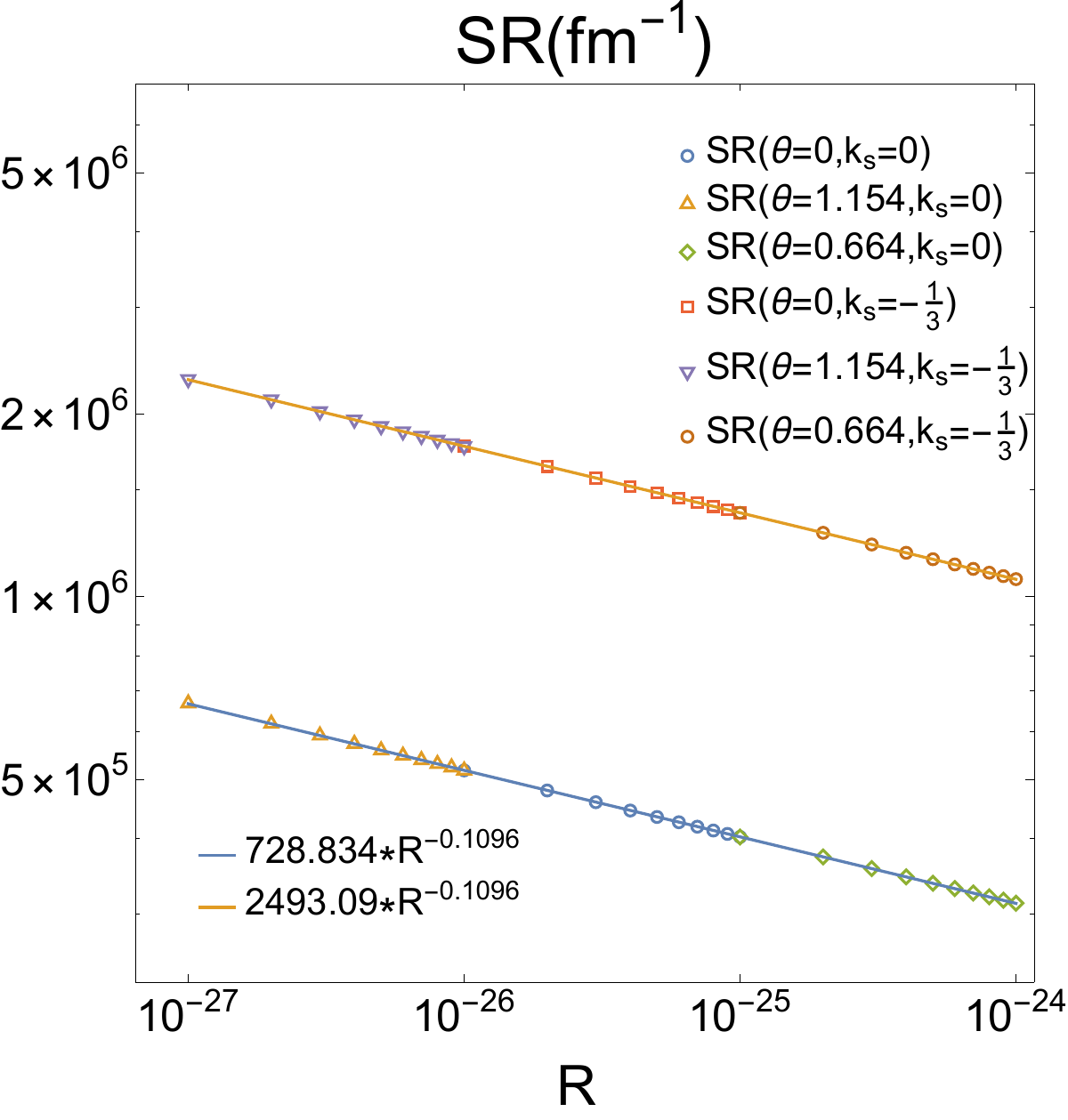}
    \caption{(Color online) The scale dependences of $\hat{q}$ (left) and $SR$ (right) with respect to $R$ for different $k_s$ and $\theta$ when approaching the CEP.}
    \label{fig:Scaling}
\end{figure}

To plot $\hat q_\sigma$ on the phase diagram, one need to know the $t$-$\bar\mu$ or $r$-$h$ dependence of the correlation length $\xi$, which is given in Ref.~\cite{ZinnJustinBook} as \begin{equation}
    \xi \approx \xi_0 R^{-\nu} \sqrt{1-\frac{5 \theta^2}{18}},
\end{equation}
where $R$ and $\theta$ are the re-organized thermal parameters expressed as
\begin{align}
\label{eq:ParameterTransfer}
&r=R(1-\theta^2),~~~
h=h_0 R^{\beta\delta}\tilde{h}(\theta),\nonumber\\
&\tilde{h}(\theta)=\theta+h_1\theta^3 +h_2\theta^5.
\end{align}
The parameters are taken from Ref.~\cite{ZinnJustinBook} as $h_0=0.394$, $h_1 =-0.76201$, and $h_2 =0.00804$. Note that when $R$ approaches $0$, both $r$ and $h$ vanish, and the system approaches the CEP. 

The jet quenching parameter $\hat q_\sigma$ and the corresponding scattering rate $SR$, obtained by removing $k_\perp^2$ in the integration Eq.~(\ref{eq:qhatSeparate}), with different values of $k_s$ are plotted in Fig.~\ref{fig:qhat and SR k0} and Fig.~\ref{fig:qhat and SR k3}. We see in both cases, $\hat{q}$ and $SR$ are enhanced on the phase boundary and diverge at the CEP, which behave exactly like the classical critical opalescence.

For investigating the divergence in detail, we plot both $\hat{q}$ and $SR$ within the extremely small $R$ region with various $\theta$ and $k_s$ in Fig.~\ref{fig:Scaling}. In all these cases, $\hat{q}$ and $SR$ exhibit the scaling behavior with almost the same critical exponent, i.e.,
\begin{equation}
    \hat{q}_{\sigma}\sim SR_{\sigma}\sim R^{-0.11}\sim R^{-\alpha}=\xi^{\frac{\alpha}{\nu}}
\end{equation}

To better understand the scaling behavior of $\hat q$ and $SR$, one can look back into Eq.~(\ref{eq:qhatSeparate}). Although it is an integration over $d^3 \mathbf k$, due to the on-shell condition that $(q+k)^2=0$, it is actually an integration over $|\mathbf k|$ and $k^0$, i.e.,
\begin{equation}
    \hat q \propto \int_0^\infty d |\mathbf k| \int_{|q-\mathbf k|-q}^{|\mathbf k|} d k^0 \cdots \widetilde{\langle\phi^\prime_{i}\phi^\prime_{j}\rangle},
\end{equation}
where $\cdots$ represents a non-singular function of $|\mathbf k|$ and $k^0$. On the other hand, when approaching the CEP,
\begin{equation}
    \begin{aligned}
    \widetilde{\langle\phi^\prime_{i}\phi^\prime_{j}\rangle}\sim
        \left[\begin{tabular}{c c}
     $\frac{2\lambda_m |\mathbf k|^2\xi^{\frac{2\alpha}{\nu}}}{\lambda_m^{2}|\mathbf k|^{4}+ k^{2}_{0}\xi^{\frac{2\alpha}{\nu}}}$ & $0$ \\
     $0$ & $\frac{\frac{T}{8\eta_{0}}\xi^{\eta}_{0}|\mathbf k|^{1+\eta}}{k^{2}_{0}+(\frac{T}{16\eta_0})^{2}|\mathbf k|^{6}}$
    \end{tabular}\right].
    \end{aligned}
\end{equation}
Notice that $\widetilde{\langle\phi^\prime_{1}\phi^\prime_{1}\rangle}$ peaks at $k^0=0$ and drops drastically when $|k^0|>\upsilon_{\mathbf k}\xi^{-\alpha/\nu}$, where $\upsilon_\mathbf k$ is an analytic function of $\mathbf k$. Therefore, the integration over $k^0$ can be estimated as
\begin{eqnarray}
    \hat q &\propto& \cdots + \int_0^\infty d |\mathbf k| \int_{-\upsilon_{\mathbf k}\xi^{-\frac{\alpha}{\nu}}}^{\upsilon_{\mathbf k}\xi^{-\frac{\alpha}{\nu}}} d k^0 \cdots \widetilde{\langle\phi^\prime_{1}\phi^\prime_{1}\rangle}\nonumber\\
    &\propto& \cdots + \int_0^\infty d |\mathbf k| \cdots \upsilon_{\mathbf k}\xi^{-\frac{\alpha}{\nu}} \widetilde{\langle\phi^\prime_{1}\phi^\prime_{1}\rangle}|_{k^0\to 0} \propto \xi^{\frac{\alpha}{\nu}},
\end{eqnarray}
which is consistent with the numerical results shown in Fig.~\ref{fig:Scaling}.  

In short, similar to the classical critical opalescence where scatterings between the light and the medium are drastically enhanced, both the $\hat q$ and the scattering rate are found divergent at the CEP with the scaling rule $\hat q \propto SR \propto \xi^{\alpha/\nu}$. This also explains why we did not see special behavior around the CEP in the previous section within the QM model with the mean field approximation, since $\alpha=0$ in all mean field theories.

{\em \color{violet} Summary. --}
We develop a new methodology for calculating $\hat{q}$ at finite chemical potential, especially in the vicinity of CEP. Based on the QM model, $\hat q$ can be decomposed into the contributions from the $\sigma$ and $\pi$ exchange processes. A perturbative calculation up to the one-loop order indicates that both contributions almost vanish in the chiral symmetry breaking (hadronic) phase, and increase not only with temperature but also with chemical potential. In the chiral symmetry restored (QGP) phase, both contributions are approximately proportional to the parton number density as $\hat{q}_{\sigma} \sim 0.2 n_{\rm parton}$, $\hat{q}_{\pi} \sim 0.6 n_{\rm parton}$, thus $\hat{q} \sim 0.8n_{\rm parton}$. The dimensionless ratio $\hat q / n_\mathrm{parton}$ is enhanced near the phase boundary, especially close to the first order phase boundary. We further evaluate $\hat q_\sigma$ in the vicinity of CEP by introducing a mapping between QCD and the Ising Model, whose critical behaviors could be in the same universality class. Based on this method, we find that both the quenching parameter and the scattering rate diverge at the critical point with the scaling behavior $\hat{q}\propto SR\propto\xi^{\frac{\alpha}{\nu}}$, and the PCO, i.e., a prominent enhancement of scatterings between jet partons and  medium via $\sigma$ exchange, is discovered near CEP. Therefore, PCO effects on jet spectra and their substructures could serve as a novel probe of CEP, with certain advantages over the traditional ones.  Our study opens a brand new perspective for improving our knowledge on the QCD phase diagram.

\acknowledgments
{\em \color{violet} Acknowledgements. --} After accomplishing this work, we became aware of other efforts on calculating $\hat q$ at finite chemical potential with fairly different conclusions~\cite{Zhu:2020qyw,Grefa:2022sav,McLaughlin:2021dlq}. We are grateful for valuable discussions with Abhijit Majumder, Xiaofeng Luo, Yi Yin, Xin-Nian Wang, Shuai Y.F. Liu and Jingyi Chao. This work is supported by the National Natural Science Foundation of China (NSFC) under Grant Nos.~12105129, 12175122, 2021-867, and  NSFC Grant No. 12247101.

\bibliography{SCrefs}


\end{document}